# Unusual kinetic properties of usual Heusler alloys

V.V. Marchenkov[1,2,*], V.Yu. Irkhin[1], A.A. Semiannikova[1,**]

[1]M.N. Mikheev Institute of Metal Physics, Ekaterinburg, Russia

[2]Ural Federal University, Ekaterinburg, Russia

*e-mail: march@imp.uran.ru

**e-mail: semiannikova@imp.uran.ru

**Abstract**

The review considers various groups of Heusler compounds, which can have the properties of a semiconductor, a half-metallic ferromagnet, a spin gapless semiconductor, a topological semimetal, and a noncollinear antiferromagnet. In these Heusler compounds, "conventional" from the point of view of the crystal structure, unusual kinetic and magnetic properties can be observed, which are caused by the features of their electronic structure (e.g., presence of an energy gap for one spin projection) and magnetic state (e.g., strong ferromagnetism, compensated ferrimagnetism, etc.). Their magnetic and kinetic characteristics are very sensitive to external influences. Depending on the alloy composition and external parameters, transitions between the considered states can be realized. All this opens up further prospects for controlling the electronic and magnetic characteristics of such compounds and their practical application.

**Introduction**

Heusler alloys were discovered by the German chemist F. Heusler as early as 1903. [1]. To date, about 1500 different Heusler compounds are known, most of which have the formula $XYZ$ (half-Heusler alloys) and $X_2YZ$ (Heusler alloys), where $X$ and $Y$ are usually transition metals, and $Z$ are $s$- and $p$-elements of the main subgroup of the Periodic Table. Fig. 1 shows the various combinations of elements that can form Heusler alloys $X_2YZ$.

**Fig. 1** Periodic table of elements. Different colors indicate the elements, the appropriate combination of which can form a Heusler alloy

For more than a hundred years, these compounds have attracted great attention and have been subjected to intensive experimental and theoretical studies, since they are of great interest both from the points of view of fundamental science and practical application. This is largely due to the fact that Heusler alloys are multifunctional materials, i.e. compounds with a combination of two or more functional properties and characteristics, such as giant magnetoresistivity [2, 3]; shape memory effect [4, 5]; large magnetocaloric effect (MCE) [6-8]; a high degree of spin polarization of current carriers (in half-metallic ferromagnets [9, 10] and spin gapless semiconductors [11, 12]); anomalous thermal properties [13, 14]; large thermoelectric effect in thermoelectrics and semiconductors [15-18]; superconductivity [19, 20]; giant anomalous Hall and Nernst effect in topological semimetals (see reviews [21, 22] and references therein), etc.

Most of the Heusler compounds have an "ordinary" crystal structure, but with features of the electronic and magnetic subsystems, leading to "unusual" kinetic and magnetic properties. As one of the first examples, one can recall the discovery by F. Heusler of alloys named after him in the $Cu_2MnAl$ compound. Three non-ferromagnetic metals, copper, manganese and aluminum, were used, as a result of which the alloy $Cu_2MnAl$ was obtained, which turned out to be a strong ferromagnet with a magnetic moment exceeding the magnetic moment of Fe. Since then, many "unusual" electronic transport and magnetic properties have been found in these "ordinary" compounds. Some of these "anomalies" can be observed in Heusler compounds exhibiting semiconducting properties, in the states of half-metallic ferromagnet (HMF) and spin gapless semiconductor (SGS), topological semimetal (TSM), and noncollinear antiferromagnet (nAFM).

To date, there are many reviews devoted to the study of the structure and physical properties of Heusler alloys (see, for example, [4, 22-28]). However, many of these reviews are devoted to describing a wide range of functional characteristics of Heusler alloys, where insufficient attention is paid to the semiconducting properties, the states of HMF and SGS, TSM, and noncollinear antiferromagnet. Therefore, the purpose of this review is to draw attention to the anomalies and unusual behavior of the kinetic properties that occur in Heusler alloys in the mentioned states, and to discuss the current state of affairs in this area.

This review will consider Heusler alloys exhibiting semiconductor properties, the properties of a half-metallic ferromagnet, a spin gapless semiconductor (Fig. 2), a topological semimetal (Fig. 3), and a noncollinear antiferromagnet (Fig. 4).

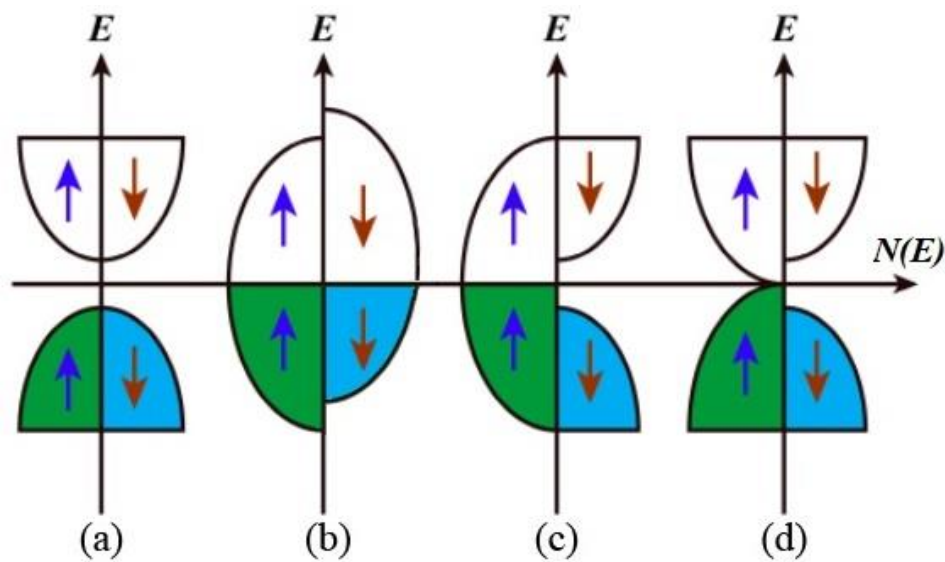

**Fig. 2** Simple models of semiconductors (a), ferromagnets (b), half-metallic ferromagnets (c), spin gapless semiconductors (d)

There is an energy gap between the top of the valence band and the bottom of the conduction band in semiconductors. The Fermi level lies within the bandgap (Fig. 2a). Semiconductors have no conduction electrons at 0 K, only when the temperature rises they appear. The number of conduction electrons grows according to the law: $exp(-\Delta E/k_B T)$, where $k_B$ is Boltzmann constant, $\Delta E$ is activation energy, that is equal to the energy gap. As the temperature rises, a certain number of electrons from the filled valence band will make quantum transitions to the empty conduction band: the intrinsic conductivity of the semiconductor arises. It should be noted that intrinsic conductivity plays a significant role only at sufficiently high temperatures; whereas at low temperatures, the origin of conduction electrons is associated with the presence of impurity atoms in the crystal [29].

Two cases of the band structure of metals are possible: if one of the bands is not completely filled, or if the bands overlap. In a metal, the Fermi surface is located inside one of the energy bands. Electrons are not localized at individual atoms, but can move freely throughout the entire volume of the metal. In the case of ferromagnetism, the resulting spontaneous magnetization is associated with the removal of spin degeneracy in the system of conduction electrons. This leads to a shift in the energy levels for electrons with "up" and "down" spin projections [29].

An interesting feature is observed in the electronic energy spectrum near the Fermi level $E_F$ in the HMF state. For one spin direction, usually "spin down", there is a wide energy gap (~ 1 eV) at $E_F$, while for the opposite direction of spin, "spin up", the band gap is absent at $E_F$ (Fig. 2c). At the same time, SGS materials possess a wide gap at $E_F$ for "spin down" electrons and a zero gap for spin up electrons, i.e. the valence and conduction bands touch each other at a point (Fig. 2d). Thus, it is possible to obtain 100% spin polarization of charge carriers [9-12]. The chemical composition, crystal structure and structural disorder strongly affect the described band structures. This applies especially to HMF- and SGS-states (see, e.g., reviews and books [12, 22-24, 27, 28] and references there).

Topological semimetals (TSMs) are characterized by bulk band crossings in their electronic structures, which are expected to give rise to gapless electronic excitations and topological features that underlie exotic physical properties. TSMs can be classified according to the dimensionality of the band crossings. Zero-dimensional (0D) band crossings are generally known as nodes or nodal points. The most famous examples include Dirac semimetals (DSMs) and Weyl semimetals (WSMs), in which two doubly or singly degenerate bands cross each other at discrete points near $E_F$, forming the fourfold Dirac points or twofold Weyl points (Fig. 3). The corresponding low-energy excitations behave as Dirac and Weyl fermions, respectively, in high-energy physics. [21].

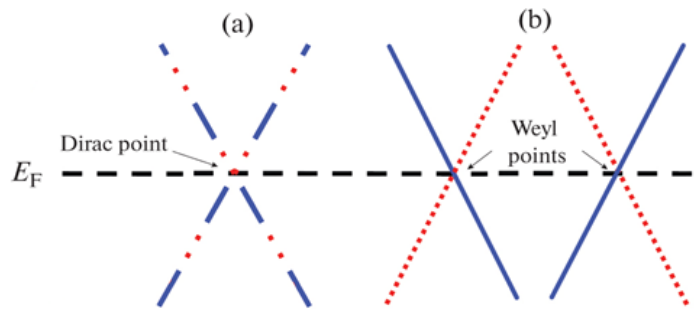

**Fig. 3** Schematic view of the band structure of (a) massless Dirac and (b) Weyl fermions. The latter arise during the decay of the Dirac point. Curves and lines consisting of solid and dashed lines represent doubly degenerate bands, and only those of solid or dashed ones represent nondegenerate bands

Symmetry breaking with respect to inversion or time reversal is the condition for the existence of a Weyl semimetal. In the bulk of Weyl semimetals there are particles with zero effective mass – Weyl fermions. Weyl nodes always appear in pairs of opposite chirality. Such nodes could be characterized as Berry curvature[1] monopoles in momentum space. In Weyl semimetals, unusual topologically protected surface states appear due to the nontrivial topology of the bulk band structure: Fermi arcs that connect the projections of the Weyl nodes onto the surface. Whereas, the Dirac nose can be considered as the "total" of two Weyl knots of opposite chirality, i.e. Dirac semimetals are not characterized by chirality. The coexistence of both symmetries with respect to inversion and time reversal in a crystal leads to the appearance of a Dirac semimetal phase [27].

Antiferromagnetic ordering is observed at low temperatures in a vast amount of crystals. Magnetic ions can be grouped into two sublattices with antiparallel average spins yielding the collinear antiferromagnetic structure [29].

The combination of noncollinear magnetic structure and Berry curvature can lead to a non-zero topological anomalous Hall effect, which is observed in $Mn_3Sn$ and $Mn_3Ge$ antiferromagnets (see [22] and references there). In addition to this Berry curvature in **k**-space, such compounds with noncollinear magnetic structures also have topological states in real space in the form of magnetic anti-skyrmions. The ability to directly control the Berry curvature shows the importance of understanding both the electronic and magnetic structure of intermetallic compounds $Mn_3X$ ($X$ = Al, Ga, Ge, Sn, etc.).

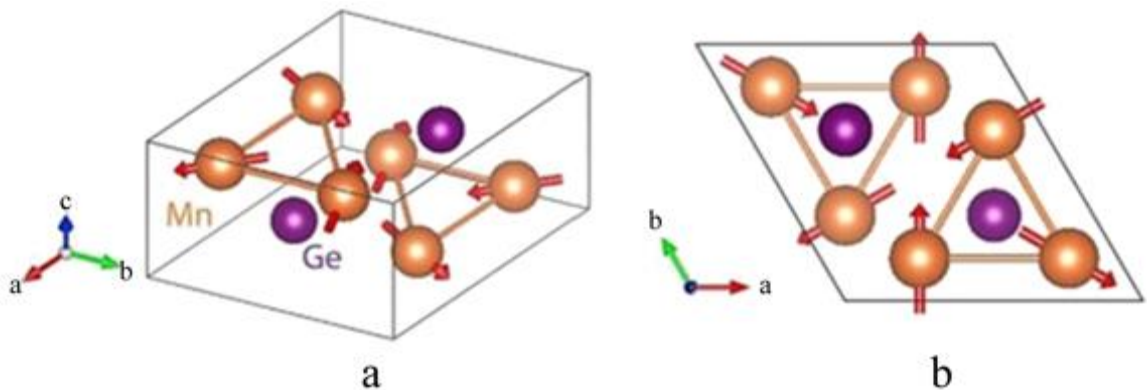

**Fig. 4** $D0_{19}$ structure of $Mn_3Ge$, which is noncollinear antiferromagnet [30]. The arrows show the direction of the magnetic moments of the Mn atoms

Compensated metallic ferrimagnets consist of inequivalent sublattices and provide an interesting alternative as they combine the high-speed advantages of antiferromagnets with those of ferromagnets, namely, the ease to manipulate their magnetic state. A compensated half-metallic ferrimagnet was first predicted in [31]. According to this model two magnetic ions in crystallographically different positions couple antiferromagnetically and perfectly compensate each-other, but only one of the two contributes to the states at the Fermi energy responsible for electronic transport [32, 33].

---

[1] Berry curvature is a topological characteristic associated with the Berry phase (which is a geometric phase difference acquired over the course of a cycle) per unit area in the quasimomentum space.

When there are two types of current carriers, e.g. with spins up and down ("majority" and "minority"), see Fig. 2 b, c, d, to describe the kinetic properties, a two-current model is used [34-36]. In this case, the system under study can be represented as two conductors connected in parallel. The conductivity of such a system is represented as the sum of the contributions from the current carriers along the magnetization (up) and against it (down):

$$\sigma = \sigma_\uparrow + \sigma_\downarrow . \qquad (1)$$

Then the resistivity has the form:

$$\rho = \sigma^{-1} = \left(\frac{1}{\rho_\uparrow} + \frac{1}{\rho_\downarrow}\right)^{-1} = \frac{\rho_\uparrow \rho_\downarrow}{\rho_\uparrow + \rho_\downarrow} \qquad (2)$$

Then it is necessary to take into account the transitions between both types of current carriers, which occur due to spin-flip scattering processes on spin inhomogeneities, magnetic impurities, etc. Extracting the contributions of these processes from the partial resistivities, we have

$$\rho_\uparrow = \rho_\uparrow^0 + \rho_{\uparrow\downarrow}, \qquad \rho_\downarrow = \rho_\downarrow^0 + \rho_{\uparrow\downarrow} \qquad (3)$$

Substituting (3) into (2), we obtain

$$\rho = \frac{\rho_\uparrow^0 \rho_\downarrow^0 + \rho_\downarrow^0 \rho_{\uparrow\downarrow} + \rho_\uparrow^0 \rho_{\downarrow\uparrow} + \rho_{\uparrow\downarrow}\rho_{\downarrow\uparrow}}{\rho_\uparrow^0 + \rho_\downarrow^0 + \rho_{\uparrow\downarrow} + \rho_{\downarrow\uparrow}} \qquad (4)$$

Such a model can be used to describe the kinetic properties in ferromagnets (Fig. 2b), HMF (Fig. 2c) and SGS (Fig. 2d). Note that for HMF and SGS at $T \ll T_c$, $T_{gap}$, basically only one conduction channel for spin up carriers operates.

**Semiconductor-like Heusler alloys**

The electrical resistivity of alloys based on $Fe_2VAl$ was studied in [16], and it was found that at $T = 2$ K, the residual resistivity $\rho_0$ of the $Fe_2VAl$ compound reaches a value of about 3000 µΩ×cm, and the temperature dependence $\rho(T)$ has a semiconductor form, i.e. greatly decreases with increasing temperature. It should be emphasized that the $Fe_2VAl$ alloy consists of "good" metals Fe, V, and Al with electrical resistivity values from a few to tens of $\mu\Omega\times$cm and a "metallic" dependence $\rho(T)$, while the "resulting" Fe$_2$VAl compound based on them is a semiconductor with the resistivity value is higher than several orders of magnitude.

Fig. 5 shows the temperature dependences of the electrical resistivity $\rho(T)$ of Fe$_{1.9}$V$_{1.1}$Al alloys [18]. It can be seen that the residual resistivity $\rho_0$ of the Fe$_{1.9}$V$_{1.1}$Al alloy reaches a huge value $\rho_0 \approx 11000$ $\mu\Omega\times$cm and demonstrates a sharp "semiconductor" decrease in resistivity with temperature. From measurements of the Hall effect [18], the concentration of current carriers $n$ was estimated. The inset to Fig. 5 shows the temperature dependence $n(T)$, which decreases with decreasing temperature, reaching the value $n \approx 5\cdot10^{19}$ cm$^{-3}$ at $T = 2$ K. In [37, 38], the electrical resistivity and the Hall effect was studied in the Fe$_2$VAl alloy, the value of the residual resistivity $\rho_0 = 2020$ $\mu\Omega\times$cm was obtained, and an estimate was made of the concentration of current carriers $n \approx 2\cdot10^{19}$ cm$^{-3}$ at $T=4.2$ K. Note that the value of the residual resistivity of the Fe$_2$VAl compound is comparable to the value of $\rho$ for semiconductors, although in terms of the concentration of current carriers $n$ is closer to the class of semimetals.

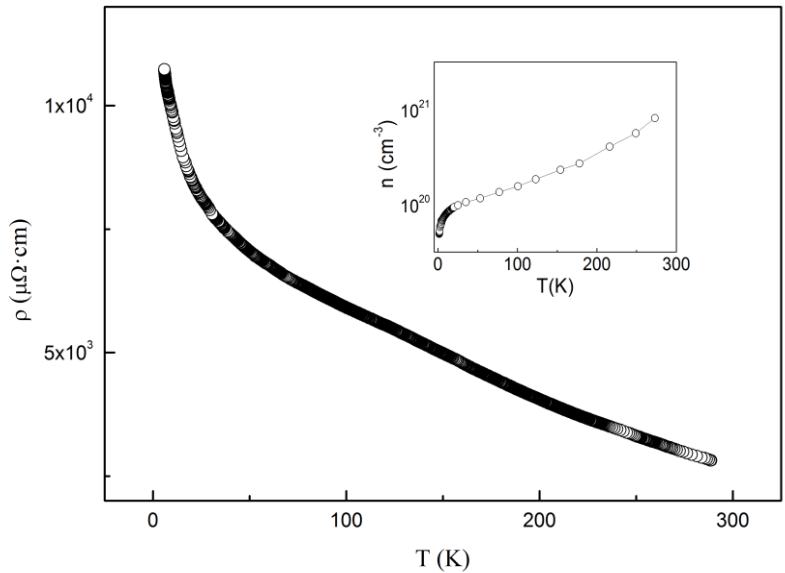

**Fig. 5** Temperature dependences of the electrical resistivity of Fe-V-Al alloys. The inset shows the temperature dependences of the charge carrier concentration [18]

The anomalous behavior of the electrical resistivity $\rho(T)$ and low values of the current carrier concentration $n$ were explained in [18, 39] by the appearance of a narrow pseudogap, i.e., a deep minimum in the density of electronic states near the Fermi level $E_F$. Fig. 6 shows the distribution of the density of electronic states of the Fe$_2$VAl Heusler alloy calculated for the ordered state [40]. It can be seen that near $E_F$, see Fig. 6, vertical dashed line, there is a gap for both spin up and spin down carriers.

It turned out that pseudogap-related anomalies are observed as well in other kinetic and magnetic characteristics of alloys based on $Fe_2VAl$ [13, 14, 41-43]. In [41], the thermoelectric power α(T) of the $Fe_{1.9}V_{1.1}Al$ compound was studied and it was found that at $T < 20$ K, a large non-monotonic contribution α(T) is observed, the presence of which can be explained by the existence of a pseudogap near the Fermi level $E_F$. The thermal conductivity and heat capacity of Heusler compounds based on $Fe_2VAl$ were studied in [13, 14, 42]. It was shown in [14] that the main contribution to the low-temperature (at $T < 15$ K) thermal conductivity κ comes from current carriers: κ(T) varies linearly with temperature, and the slope tangent of κ(T) is largely related to the pseudogap near $E_F$. When studying the low-temperature part of the heat capacity of alloys based on $Fe_2VAl$ [13, 14, 42, 44], direct experimental evidence was obtained for the existence of such a pseudogap, the value of which, according to the estimates made, is about 2 meV.

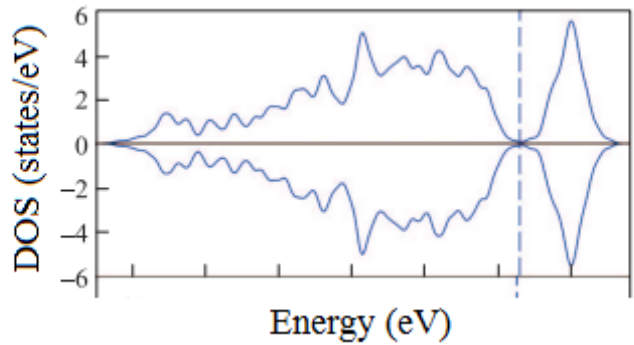

**Fig. 6** The distribution of the density of electronic states of the Heusler alloy $Fe_2VAl$, calculated for the ordered state [40]

Thus, Heusler alloys based on $Fe_2VAl$, which are prepared from "good" metals, exhibit unusual "semiconductor" low-temperature kinetic and thermal properties. Their unusual behavior can be explained by the manifestation of a narrow pseudogap in the density of electronic states at $E_F$.

**Half-metallic ferromagnetic Heusler alloys**

Half-metallic ferromagnets were theoretically predicted in 1983 by R. de Groot, who performed band calculations for the NiMnSb compound [45]. The main feature of HMF is the presence of a gap at the Fermi level for electronic states with spin down, see Fig. 2c. As mentioned above, to describe the kinetic properties of materials with very different states for opposite spin projections, primarily electrical conductivity, the two-current model of conductivity can be used [34-36]. In this case, the total resistivity of such a system is described by the formula (4). At temperatures much lower than the gap in HMF, only one of the conducting channels operates, namely, the channel for spin-up carriers. Since such electronic states are "metallic" (Fig. 2c), metallic properties should appear under these conditions. For electrical resistivity, these are small values of residual resistivity and a metallic type of conductivity.

Many Heusler alloys based on cobalt $Co_2YZ$ ($Y$ = Ti, V, Cr, Mn, Fe, Co, Ni; $Z$ = Al, Ga, Ge, Si, Sn, In, Sb) have low residual resistivity and positive temperature coefficient of resistivity [28, 46-48]. Such alloys, in particular, include $Co_2FeSi$ and $Co_2MnSi$ compounds. Fig. 7 shows the density of electronic states of the Heusler alloy $Co_2MnSi$, the band gap having value 0.77 eV [49].

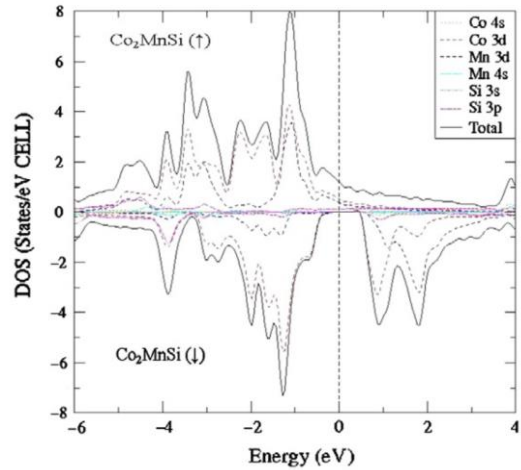

**Fig. 7** The total and partial DOS of $Co_2MnSi$ Heusler alloys calculated by Generalized Gradient Approximation (GGA) method [49]

It can be seen that for spin down electronic states there is a gap near $E_F$, which is absent for spin up states. A similar pattern is observed for other compounds of the $Co_2MnZ$ system and $Co_2FeSi$ as Figs 7 and 8 show.

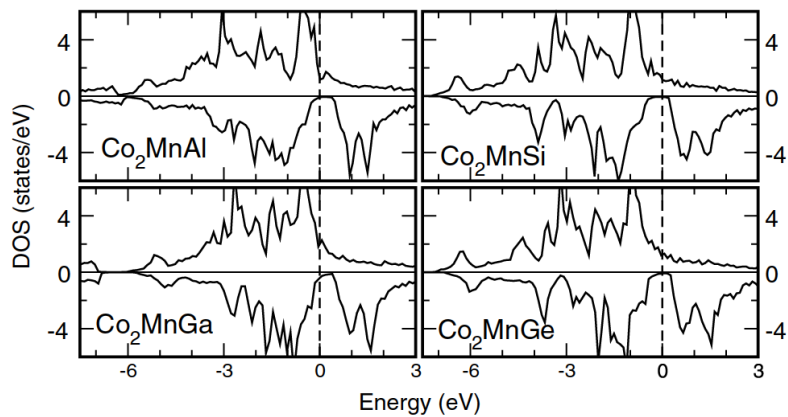

**Fig. 8** Atom-resolved DOS for the $Co_2MnZ$ compounds with $Z$ = Al, Si, Ge, Ga elements calculated using the full-potential screened Korringa-Kohn-Rostoker (FSKKR) method [50]

In [51], the authors synthesized a Co$_2$FeSi single crystal and studied the electrical resistivity and galvanomagnetic properties in the temperature range from 4.2 K to 300 K. The authors of [51] showed that the residual resistivity of this crystal is quite small and $\rho_0 = 4.3$ µΩ×cm, and $\rho(T)$ increases with increasing $T$ (Fig. 9). In this case, the temperature-dependent part of the resistivity can be represented as the sum of the contributions from the electron-phonon $\rho_P$ and electron-magnon $\rho_M$ scattering mechanisms, and $\rho_M$ can be written in the form:

$$\rho_M(T) = cT^2 \times e^{-\frac{\Delta}{T}}, \qquad (5)$$

where $c$ is efficiency parameter of electron-magnon scattering, $\Delta$ is gap size. Using the fitting, the authors determined the size of the gap, $\Delta = 103$ K, which corresponds to the energy $k_B\Delta = 8.9$ meV. As can be seen, the experimentally determined size of the gap is an order of magnitude smaller than theoretical calculations give [49].

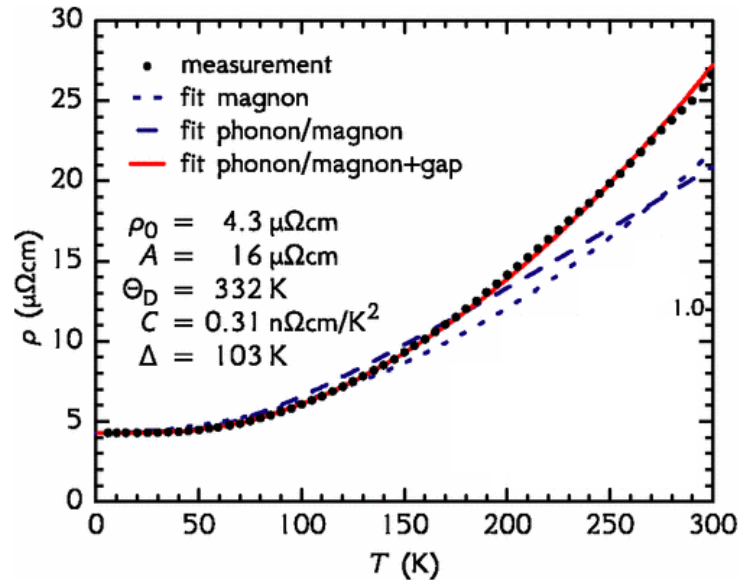

**Fig. 9** The temperature dependence of the electrical resistivity of Co$_2$FeSi. Black dots mean measured values. Fits with magnonic and magnonic-phononic temperature dependence are displayed as blue dashed lines. The red continuous line shows a fit to the electron-magnon scattering with an additional Boltzmann factor [51]

Knowing the size of the gap Co$_2$FeSi, $\Delta_{Co2FeSi} = 0.1$ eV, the corresponding temperature can be estimated, $T_{Co2FeSi} \approx 1100$ K [51], which is comparable to the Curie temperature of Co$_2$FeSi compound, $T_C = 1100$ K [52]. Then at low temperatures, up to room temperature, only one conduction channel for current carriers with spin up should work. Therefore, the use of an exponential term to describe the temperature dependence of the resistivity under these conditions is hardly justified. Under conditions where there is one type of current carriers, a high degree of polarization of current carriers can occur, and specific mechanisms of scattering of spin-polarized carriers appear in the kinetic properties, in particular, in electrical resistivity.

Mechanisms of two-magnon scattering leading to a power-law temperature dependence of the resistivity $\rho(T) \sim T^n$, where the exponent $n = 9/2$ at $T > T^{**}$ and $n = 7/2$ at $T > T^{**}$, with temperature $T^{**} \sim q_2^2 T_C$, were predicted in [53]. Negative linear magnetoresistivity can also be a manifestation of the process of two-magnon scattering.

In a simple single band model of HMF, where $E_F < \Delta$, $q_2 \sim (\Delta/W)^{1/2}$ ($W$ is bandwidth). It is worth noting that, $q_2$ can be quite small, provided that the band gap is much smaller $W$, which is typical for real HMF systems [10, 53].

Predicted in [53] the dependence $\rho(T) \sim T^{9/2}$ was experimentally observed in [54], where the temperature dependences of the electrical resistivity of Heusler alloys based on $Co_2FeSi$ with the replacement of half of the Si atoms by Al, Ga, Ge, In and Sn were studied. It was found that at low temperatures up to 50 – 80 K, these dependences $\rho(T)$ can be described by the formula

$$\rho(T) = \rho_0 + aT^2 + bT^{9/2}, \qquad (6)$$

where $\rho_0$ is residual resistivity.

The temperature dependences of the resistivity of a $Co_2FeSi$ single crystal were measured in magnetic fields from 0 to 150 kOe in [55]. There are three temperature intervals in which the resistivity depends differently on temperature and magnetic field (Fig. 10). Firstly, below 30 K $\rho(T) \propto AT^n$, $n \approx 2$ and coefficient $A \propto H^2$. Secondly, from 30 to 60 K $\rho(T) \propto CT^n$, $n \approx 4$ and coefficient $C \propto -H$. Thirdly, above 65 K $\rho(T) \propto BT^n$, $n \approx 2$ and coefficient $B \propto H^{-2}$. Therefore, a manifestation of two-magnon scattering processes that determine the behavior of the electrical and magnetoresistivity of the alloy in the temperature range 30 K $< T <$ 60 K can be a power-law temperature dependence of the electrical resistivity with the highest exponent and linear negative magnetoresistivity.

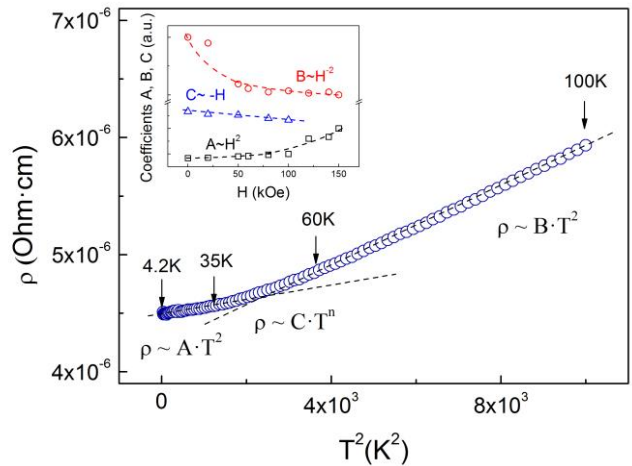

**Fig. 10** Dependences of the electrical resistivity of the $Co_2FeSi$ alloy on the squared temperature. The inset shows the field dependences of the coefficients *A*, *B*, and *C* [55]

Under these conditions, a high degree of polarization of current carriers can occur with a polarization coefficient $P$ close to 1. The spin polarization coefficient $P$ is given by the following expression:

$$P = (N_\uparrow(E_F) - N_\downarrow(E_F))/(N_\uparrow(E_F) + N_\downarrow(E_F)), \tag{7}$$

where $N_\uparrow(E_F)$ and $N_\downarrow(E_F)$ are densities of electronic states at the Fermi level $E_F$ with the spin up and spin down, respectively. Then, in a simple mean-field approximation, the presence of only one type of carriers can mean 100% spin polarization of current carriers, which somewhat decreases when non-quasiparticle states are taken into account [10, 56, 57].

Epitaxial films of the Heusler alloy $Co_2MnSi$ were synthesized in [46] and their characteristics were studied by ultraviolet-photoemission spectroscopy, in particular, the polarization coefficient of current carriers $P$ at room temperature was determined. Authors of [46] experimentally showed that $P$ (×100) = $93^{+7}_{-11}$. Experimental studies of the kinetic and magnetic properties of Heusler alloys based on $Co_2YZ$ ($Y$ = Ti, V, Cr, Mn, Fe, Co, Ni; $Z$ = Al, Ga, Ge, Si, Sn, In, Sb) demonstrates that residual resistivity of such compounds is quite small (from 1.6 to 50 $\mu\Omega\times$cm ), dependence $\rho(T)$ possesses a metallic type, the concentration of current carriers also corresponds in magnitude to the "metallic" value, i.e. ~ $10^{22} - 10^{23}$ cm$^{-3}$, the magnetic state, as a rule, is ferromagnetic with sufficiently large magnetization values up to 6 $\mu_B$/f.u. [48, 58, 59].

**Spin gapless semiconductors based on Heusler alloys**

Recently, new "gapless" materials have been discovered, such as graphene [60], topological insulators [61] and semimetals [62]. They have unusual magnetic and electronic characteristics that are very sensitive to external influences, which may be important for practical applications. Using magnetic and electric fields, pressure and temperature change, it is possible to "tune" their electronic band structure, and, consequently, purposefully influence their transport and magnetic properties. There is no energy gap in such materials, i.e. there is no energy threshold for the transition of electrons from the valence band to the conduction band, and charge carriers in such materials, as a rule, have a large mean free path and high mobility. Similar properties are also observed in spin gapless semiconductors (SGS), a new class of materials theoretically predicted in 2008 [11]. SGS has an unusual band structure: near the Fermi level, there is an energy gap for the electron subsystem with spin down, and for current carriers with spin up, the top of the valence band touches the bottom of the conduction band (Fig. 2d). In many ways, this is similar to classical gapless semiconductors [63].

The two-current conductivity model is applicable to the SGS, by analogy with the HMF [64], and at temperatures much lower than the gap size in the SGS, only one of the conducting channels also works – the channel for carriers with spin up, see Fig. 2d. It can be assumed that, under these conditions, the kinetic properties of SGS materials should exhibit the properties of classical gapless semiconductors: a relatively large residual resistivity and a weak dependence of $\rho(T)$; a relatively low value of the current carrier concentration and a small thermoelectric power. Since in this case current carriers with spin up predominate, a high degree of spin polarization, a relatively large amount of magnetization and an anomalous Hall effect should be observed.

The SGS state with unusual magnetic and magneto-transport properties was experimentally observed in $Mn_2CoAl$ [65] and $Ti_2MnAl$ Heusler alloys [66]. In addition, it was predicted in works on calculations of the electronic band structure that the SGS state should also be observed in a number of quaternary alloys of the type CoFeMnSi, CoFeCrAl, CoMnCrSi, CoFeVSi and FeMnCrSb [67].

The authors of [65, 68] reported the first experimental observation of the SGS state in the inverse Heusler compound. The Curie temperature of this compound is $T_C = 720$ K, and the magnetic moment to be $2\mu_B$/f.u. In this case, the value of the residual resistivity is $\rho_0 = 446$ μΩ×cm at $T = 4.2$ K, which is two orders of magnitude greater than $\rho_0$ of the HMF $Co_2FeSi$ [51], and the electrical resistivity $\rho$ slightly decreases with increasing a temperature (see Fig. 11). The Seebeck coefficient is close to zero and practically has no dependence on temperature, the concentration of current carriers is relatively low and changes slightly with increasing $T$ from $1.7 \times 10^{20}$ to $3 \times 10^{20}$ cm$^{-3}$ (Fig. 11). The value of the anomalous Hall conductivity $\sigma_{xy}$ is relatively small and amounts to $\sigma_{xy} = 21.8$ S×cm$^{-1}$ at $T = 2$ K in a field of $1.6\times10^6$ A/m, i.e. around 20 kOe (Fig. 11), which is explained by the Berry curvature symmetry properties [65].

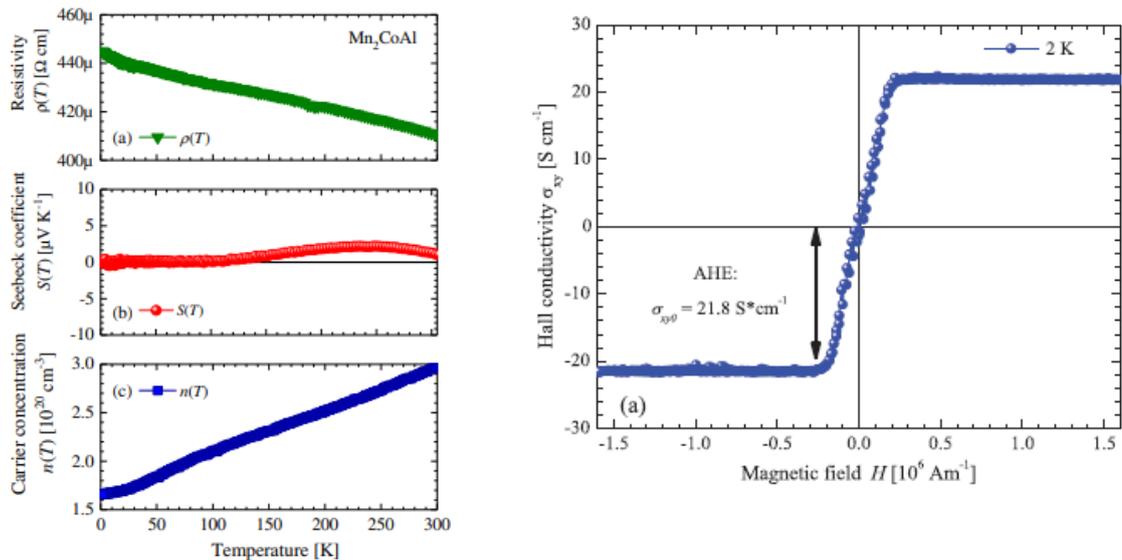

**Fig. 11** Left: temperature dependences of resistivity, Seebeck coefficient and current carriers concentration for $Mn_2CoAl$ [68]; right: field dependence of Hall conductivity for $Mn_2CoAl$ at 2 K [65]

Calculations of the electronic structure and properties, as well as the magnetic characteristics of Mn$_2$CoAl and Mn$_2$CoGa alloys, were performed in [69]. Mn$_2$CoAl and Mn$_2$CoGa alloys are SGS and HMF compounds, respectively. Therefore, the authors studied the changes in the electronic structure and magnetic state that occur in these alloys when Al and Ga atoms are replaced by Cr and Fe. It turned out that such doping leads to degradation of the SGS state of the Mn$_2$CoAl compound, since delocalized states appear in the spin-down forbidden energy region. At the same time, the HMF nature of the Mn$_2$CoGa compound does not change when it is relatively lightly doped. In this case, the magnetic moments of doped alloys increased, since nonmagnetic Al and Ga atoms were partially replaced by magnetic Cr and Fe.

In [70], thin films of the Heusler compound Mn$_{2-x}$Co$_{1+x}$Al ($0 \leq x \leq 1.75$) were synthesized and their magnetic and galvanomagnetic properties were measured. It was found that the films are ferromagnetic at $x = 1.75; 1.5; 1.25; 1$ and ferrimagnetic at $x = 0$; 0.5 and 0.75. The value of the residual resistivity in Mn$_2$CoAl films is very high and reaches a value of more than 1500 μΩ×cm, and the form of the dependence $\rho(T)$ is semiconductor-type one. The value of the anomalous Hall conductivity is about 3.4 S/cm at 10 K in a field of 64 kOe (6.4 T), and the magnetoresistivity is positive and linear in the magnetic field. The authors conclude that the SGS state appears in these films at a high Mn concentration, when $x$ is small, and such films can be used in spintronics.

Another compound exhibiting SGS properties is the quaternary CoFeCrGa Heusler alloy. The authors of [71] followed the change in its magnetic state, electronic structure, and electron transport upon partial replacement of Fe atoms by Co, i.e., when $x$ changes in Co$_{1+x}$Fe$_{1-x}$CrGa ($0 \leq x \leq 0.5$). It has been shown that the Curie temperature and magnetic moment increase from 690 K to 870 K and from 2.1 $\mu_B$/f.u. up to 2.5 $\mu_B$/f.u., as $x$ increases from 0 to 0.5, respectively. Simultaneously, the value of the residual resistivity of alloys with $x \leq 0.4$ is quite large and varies approximately from 320 to 450 μΩ×cm and slightly decreases with increasing temperature. This testifies in favor of the SGS state, as well as the value of the Seebeck coefficient close to zero. In addition, in these alloys, a fairly large anomalous Hall conductivity is observed, equal to 38 S/cm ($x = 0.1$) and 43 S/cm ($x = 0.3$) at 5 K in a field of 50 kOe, which weakly depends on temperature, decreasing by only 12 – 15% as the temperature rises to $T = 300$ K.

Another quaternary Heusler alloy CoFeMnSi, in which the SGS state can appear, was synthesized in [72]. The authors studied the structure, magnetic and electronic transport properties, Andreev reflection, and also performed calculations of the band structure. The CoFeMnSi compound was found to have a Heusler structure (of the LiMgPdSn type) with a slight disorder present. Its Curie temperature is 620 K and its magnetic moment is 3.7 $\mu_B$/f.u. The residual resistivity is relatively large and amounts to about 340 μΩ×cm. The temperature dependence of the resistivity is very weak: in the temperature range from 5 to 300 K, the value of $\rho$ decreases by only 2%. The current carrier concentration is relatively low and varies from 4.53×10$^{19}$ at 5 K to 4.85×10$^{19}$ cm$^{-3}$ at room temperature. In this case, the anomalous Hall conductivity is rather high and amounts to 162 S/cm at 5 K in a field of 50 kOe. The Andreev reflection measurements made it possible to estimate the spin polarization coefficient *P*, which at low temperatures turned out to be quite large, $P = 0.64$. All this indicates the realization of the SGS state in the CoFeMnSi compound.

The presence of the SGS state can lead to peculiarities not only in kinetic properties, but manifest itself as well in other electronic characteristics, in particular, in optical ones. In [73], the optical properties were studied and the electronic structure of the SGS compound $Mn_2CoAl$ was calculated. It was found that anomalies are observed in its optical properties, indicating a weakening of the metallic properties: positive values of the real part of the permittivity $\varepsilon_1$ and the absence of the Drude contribution to the optical conductivity in the IR region of the spectrum up to the boundary of the interval studied. It was shown that intense interband absorption in the IR region indicates a complex structure of the band spectrum and a high density of *d* states in the vicinity of the Fermi level $E_F$. The observed features of the optical absorption spectrum made it possible to explain the pattern of the band spectrum characteristic of spin gapless semiconductors.

**Topological semimetallic Heusler compounds**

The review [22] considers the electronic structure, magnetic and electronic transport properties of various Heusler compounds based on heavy elements, in which there is a strong spin-orbit interaction. It has been demonstrated that due to the symmetry features of the spin-orbit interaction and the magnetic structure in such compounds, Berry curvature arises, and this in turn leads to the appearance of numerous topological phases, including topological insulators and topological semimetals. In topological semimetals, Weyl points and node lines can appear, which can be tune by various external influences. As a result, numerous exotic properties can arise, in particular, chirality anomalies, large anomalous topological and spin Hall effects.

The electronic structure and magnetic state of the $Co_2MnGa$ single crystal using ARPES spectroscopy, density functional method were investigated in [74] where also electrical resistivity, Hall effect and magnetization were experimentally studied. It was found that $Co_2MnGa$ is a topological semimetal in which the lines of topological nodes cross the Fermi level, leading to the appearance of surface states. A large anomalous Hall conductivity was found in the $Co_2MnGa$ single crystal (Fig. 12), which can be explained by the existence of lines of topological modes. The results can be used to obtain strongly spin-polarized current carriers.

The single crystal $Co_2MnGa$ compound was studied furthermore in [75]. At room temperature in a magnetic field of 1 T, a large anomalous Nernst effect was detected, the thermoelectric power is about 6 μV/K, which is almost an order of magnitude greater than the thermoelectric power ever observed in ferromagnets. The occurrence of the detected anomaly can be explained by the large Berry curvature near the Fermi level, associated with the node lines and Weyl points.

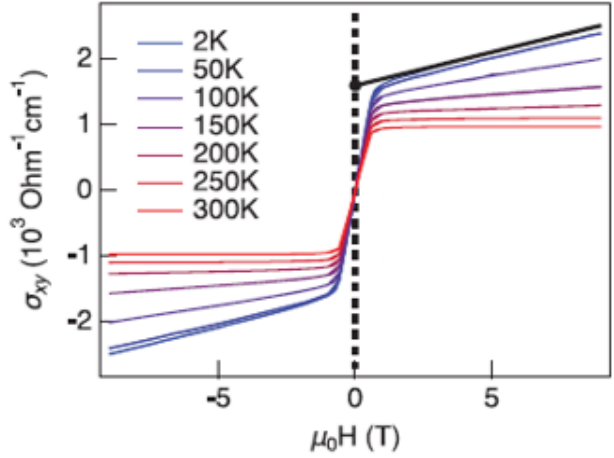

**Fig. 12** Field dependence of Hall conductivity for $Co_2MnGa$ single crystal at different temperatures [74]

Calculations of the electronic structure of the Heusler compound $Ti_2MnAl$ were carried out in [76]. The $Ti_2MnAl$ alloy is a candidate for a topological Weyl semimetal; where ferrimagnetic ordering with a Curie temperature of about 650 K is observed. Despite such a ferrimagnetic state with a near-zero magnetization value, a large anomalous Hall conductivity of about 300 S/cm is observed despite the vanishing net magnetic moment. The realization of the TSM state is explained as a consequence of the redistribution of the Berry curvature from the Weyl points, which are only 14 meV from the Fermi level and are isolated from the trivial bands [76]. It is noted that the anomalous Hall effect in $Ti_2MnAl$ arises directly from the Weyl points and is topologically protected. This is a fundamental difference from the situation with antiferromagnetic $Mn_3X$ alloys ($X$ = Ge, Sn, Ga, Ir, Rh, and Pt), where the anomalous Hall effect arises due to the noncollinear magnetic structure. Apparently, this is one of the first observations of a topological Weyl semimetal based on a Heusler alloy with a zero magnetic moment but a large anomalous Hall effect.

Calculations of the electronic structure, measurements of the electrical and magnetoresistivity, the Hall effect, and magnetization were carried out on epitaxial films of the $Co_2TiSn$ compound [77], which is considered to be a TSM with Weyl points. The Berry curvature arising in momentum space can lead to anomalies in electron transport, in particular, to a large anomalous Hall effect. Some of the $Co_2TiSn$ films had a residual resistivity of about 190 μΩ×cm, which slightly increased with temperature up to 200 μΩ×cm at room temperature. According to the data obtained, the current carrier concentration at 2 K was $1.41 \times 10^{22}$ cm$^{-3}$ and increased linearly with temperature up to $1.57 \times 10^{22}$ cm$^{-3}$ at 300 K. The absolute value of the calculated Hall conductivity reached 250 S/cm. As the authors show [77], the intrinsic contribution to the anomalous Hall effect can arise due to the node lines partially broken due to the symmetry reduction caused by the absence of time-reversal symmetry.

In [78], the anomalous Hall effect in the Heusler compounds $Co_2TiSi$ and $Co_2TiGe$ was experimentally studied in order to elucidate the role of the Berry curvature in its formation. The residual resistivity was about 110 μΩ×cm for $Co_2TiSi$ and about 142 μΩ×cm for $Co_2TiGe$. The estimated carrier density for $Co_2TiSi$ and $Co_2TiGe$ are $\sim 5\times 10^{20}$ and $\sim 6\times 10^{21}$ cm$^{-3}$, respectively, which changed slightly with *T*. It was demonstrated that the anomalous Hall resistivity, depending on the electrical resistivity, varies according to a law close to quadratic for both compounds. An analysis of the obtained data showed that the contribution to the anomalous Hall conductivity from the internal Karplus-Luttinger mechanism associated with the Berry phase prevails over the external contributions from the mechanisms of asymmetric skew scattering and side jumps.

The inverse spin Hall effect was measured in the $Co_2MnGa$ Heusler alloy, which is a ferromagnetic Weyl TSM [79]. The authors applied the spin injection method in "spin valve" type structures. The angle $\theta_{SH}$ of the spin Hall effect was determined, $\theta_{SH}$ = -0,19 ± 0,04. It turned out that its value is one of the highest for ferromagnets. In this case, the Onsager principle is failed, which may be due to the difference in the values of the Hall conductivity for current carriers with spin up and spin down.

The authors of [80] studied ferromagnetic Weyl TSM films $Co_2MnGa$ using ARPES spectroscopy, electronic structure calculations, and measurements of electron transport and magnetic properties. A spin-polarized Weyl cone was visualized, which made it possible to observe non-dispersive surface states in the region of room temperatures. It was found that under these conditions the anomalous Hall and Nernst effects increase as the magnetization-induced massive Weyl cone approaches the Fermi level. This occurs until the anomalous Nernst thermoelectric power reaches 6.2 μV/K. An unambiguous relationship was found between the topological quantum state and remanent magnetization. On this basis, the authors of [80] concluded that $Co_2MnGa$ Heusler alloys can be used to create sensors for measuring heat flux and magnetic field in the room temperature region.

In [81], the Heusler compound $Co_2MnGe$ was studied using ARPES spectroscopy. A broad band with parabolic dispersion was found at the center of the Brillouin zone, and several main spin bands with high spin polarization crossing the Fermi level near its boundary were detected. The obtained data indicate the absence of the contribution of minority spin bands on the Fermi surface and, consequently, the appearance of the TSM state in $Co_2MnGe$, which should lead to large anomalous Hall and Nernst effects.

Thus, although the study of topological phenomena in Heusler alloys is just beginning, it seems to be very promising, since, in combination with a large spontaneous magnetization and a large anomalous Hall conductivity, a number of physically interesting and practically important effects can be realized.

**Noncollinear antiferromagnets based on Mn$_3$*Z* Heusler alloys**

Noncollinear antiferromagnetism can occur in compounds with a triangular lattice [82], in particular, in alloys with a kagome-type lattice (Fig. 4). It was shown in [30, 83] that in Mn$_3$Ge and Mn$_3$Sn compounds with a distorted $D0_{19}$ Heusler structure, a triangular-type lattice of Mn atoms, a highly frustrated kagome lattice, is formed in the *ab* plane, and antiferromagnetic ordering is observed, which can lead to a number of unusual effects.

For intermetallic Mn$_3$Z (Z = Ge, Sn, Ga, Ir, Rh) compounds in the antiferromagnetic state, a strong anisotropic anomalous Hall effect and the spin Hall effect were found [84]. In [85], and zero magnetic moment was reported in thin films of the Mn$_3$Al alloy, which was explained by the state of a compensated ferrimagnet (CFiM). This state differs from antiferromagnetism (AFM) because the crystallographic positions of manganese are different. In [86], mention is made of a zero magnetic moment in the molten Mn$_3$Al alloy and it is suggested to be a manifestation of compensated ferrimagnetism.

The combination of non-collinear magnetic structure and Berry curvature may lead to a non-zero topological anomalous Hall effect, which was first observed in antiferromagnets Mn$_3$Sn and Mn$_3$Ge [22]. In addition to this Berry curvature in *k*-space, such compounds with non-collinear magnetic structures have topological states in real space as well in the form of magnetic anti-skyrmions. The ability to directly tune the Berry curvature shows the importance of understanding both the electronic and magnetic structure of compounds Mn$_3$X (X = Al, Ga, Ge, Sn, etc.).

Recently, there have been reported of new antiferromagnetic materials with a high response speed, low power consumption, and high noise immunity to an external magnetic field, which can be used in the field of spintronics [87–99]. Such antiferromagnets can be divided into two types: collinear and noncollinear ones. Traditional collinear antiferromagnetic materials have two spin sublattices with an antiparallel direction of magnetization, so that there is no macroscopic magnetic moment. In noncollinear antiferromagnetic materials, the magnetic moments of atoms are not directed strictly along one line. These are compounds such as Mn$_3$X (X = Sn, Ge, Ga), in which the Mn atoms form a hexagonal sublattice, and the magnetic moments form a kagome lattice and, accordingly, lead to the appearance of a noncollinear antiferromagnetic state. In such antiferromagnets, a large anomalous Hall effect, a new spin Hall effect, a large spin Nernst effect, and other unusual phenomena have been discovered [100–102]. Due to their rich and unusual electronic structure, as well as their magnetic order, noncollinear antiferromagnets are of significant fundamental interest in the field of spin-orbit torque research and have great potential for practical applications in spintronics.

In [103], *ab initio* calculations were carried out for a number of noncollinear antiferromagnetic materials and it was found that in the Mn$_3$Sn compound, due to the noncollinear arrangement of spins, structurally induced symmetry breaking leads to a spin current, which differs significantly from that in traditional heavy metals with large atomic numbers and strong spin-orbit coupling under spin Hall effect conditions. In this case, the direction of polarization in the spin current has components parallel to the charge current and perpendicular to the film surface.

In [104], it was experimentally demonstrated that the noncollinear antiferromagnetic material Mn$_3$Sn has more bright characteristics of the spin Hall effect, and the noncollinear structure of spins can cause a momentum-dependent spin splitting. Such spin splitting can lead to a new magnetic spin Hall effect. The authors of [105] managed to generate a spin torque using the spin Hall effect in traditional nonmagnetic heavy metals. This led to a vertical change in the antiferromagnetic domains in $Mn_3Sn$ and was successfully used to tune the spin-torque effect with electric current. In [106], a noncollinear antiferromagnetic material $Mn_3GaN$ was used as a source of spins to control the direction of spin polarization in the spin current. The presence of spin polarization in such a system was experimentally proven (Fig. 13).

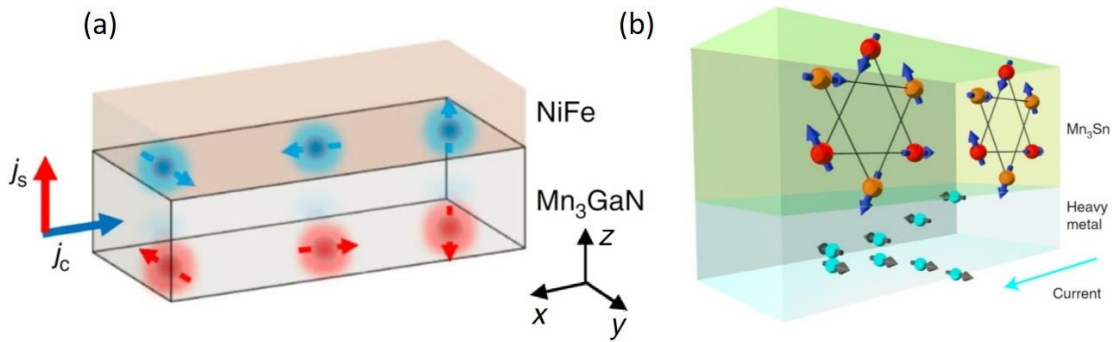

**Fig. 13** Spin-orbit torque study based on non-collinear antiferromagnetism. **(a)** A non-collinear antiferromagnetic $Mn_3GaN$/magnetic NiFe alloy material, in which the $Mn_3GaN$ material acts as a spin source layer, generates a spin current $j_S$ through a charge current $j_C$, where the spin polarization directions (respectively indicated by highlighted arrows) exist along with x, y and z-direction components [106]. **(b)** Non-magnetic heavy metal Pt/non-collinear antiferromagnetic $Mn_3Sn$, where Pt is the spin source layer generating spin current to manipulate the spin structure in $Mn_3Sn$ [107]. The dark blue arrows in the $Mn_3Sn$ layer represent the magnetic moments of Mn atoms, and the gray arrows in the heavy metal layer represent the spin polarization direction in the spin current

Based on the noncolinear antiferromagnetic material $Mn_3SnN$, the authors of [108] managed to implement switching of the perpendicular magnetic moment in the $Mn_3SnN$/(Co/Pd)$_3$ nanostructure using a spin polarized in the direction perpendicular to the film plane. In [109], antiferromagnetic $Mn_3Sn$ was used as a spin source layer, and a spin, being polarized in the direction of the vertical plane of the film, was effectively used to control the switching of the magnetic moment in Co/Ni multilayers with perpendicular magnetic anisotropy.

At present, there are a number of works on the study of the effect of the spin-orbital torque in noncollinear antiferromagnet/magnet heterostructures. The authors of [107] applied spin-orbit torque to control the spin structure in the $Mn_3Sn$ antiferromagnet, finding that the noncollinear antiferromagnetic arrangement of spins plays an important role in the spin-orbit torque effect. The new spin structure in noncollinear antiferromagnetic materials can be used to implement efficient control of the spin torque transfer in spintronic devices.

**Conclusions**

To date, about 1500 different Heusler compounds are known, many of which exhibit unusual properties in electron transport and magnetic state. As already F. Heusler discovered, the "melting" of three non-ferromagnetic metals Cu, Mn and Al into the $Cu_2MnAl$ compound leads to an unexpected magnetic state: the resulting $Cu_2MnAl$ alloy is a strong ferromagnet. In this review, we have tried to draw attention to some of the unusual kinetic and magnetic properties, as well as the reasons for their occurrence, in seemingly ordinary Heusler alloys. These are semiconductor alloys, compounds in the states of HMF, SGS, and TSM, and, finally, noncollinear AFM alloys based on $Mn_3Z$.

A semiconductor-like alloy based on $Fe_2VAl$ consists of three good metals Fe, V and Al, which have a "metallic" resistivity from a few to tens of $\mu\Omega \times cm$ and a "metallic" conduction electron concentration of ~ $10^{22} - 10^{23}$ $cm^{-3}$. While the resulting $Fe_2VAl$ Heusler alloy is semiconductor with a residual resistivity of up to 11000 $\mu\Omega \times cm$ and a current carrier concentration of ~ $10^{19} - 10^{20}$ $cm^{-3}$. The reason for this is the presence of a wide pseudogap at the Fermi level.

Heusler half-metallic ferromagnetic alloys such as $Co_2FeSi$ and $Co_2MnSi$ have high Curie temperatures and relatively large gaps for spin down current carriers. Therefore, at room temperatures and below, charge carriers with spin up are mainly involved in the conduction processes. The presence of "conducting" electrons leads to metallic conductivity with a high degree of their spin polarization, and in more complex kinetic effects, the contributions of two-magnon scattering processes and non-quasiparticle states appear [10]. As a result, a specific temperature dependence of the electrical resistivity $\sim T^n$ is observed, with an exponent $n$, $7/2 \leq n \leq 9/2$, and a negative linear magnetoresistivity. The residual resistivity in this case is quite small in magnitude, the concentration of conduction electrons corresponds to the "metallic" ~ $10^{22} - 10^{23}$ $cm^{-3}$, the magnetic moment corresponds to a good ferromagnet, the Hall effect is relatively small.

By analogy with HMF, for spin gapless semiconductors at temperatures much lower than the gap energy, only one conduction channel also operates – for current carriers with spin up. Since such carriers are in states characteristic of gapless semiconductors, the kinetic properties under these conditions are mainly formed by such carriers. Therefore, SGS materials, one of the brightest representatives of which is the $Mn_2CoAl$ compound, are characterized by: 1) high values of the residual resistivity ~ 400 $\mu\Omega \times cm$ and a relatively low concentration of current carriers ~ $10^{20}$ $cm^{-3}$; 2) weak dependence of electrical resistivity on temperature; 3) almost zero thermoelectric power; 4) large values of the magnetic moment; 5) large anomalous Hall effect.

Heusler compounds belonging to the class of topological systems exhibit exotic properties, in particular, of various types of topological Weyl semimetals. Such materials are characterized by a large residual resistivity and its weak dependence on temperature, a relatively low concentration of current carriers, a large anomalous Hall effect, and a large thermoelectric power. Such Heusler compounds are quite promising as model systems that can be used for model calculations and experiments in order to study and understand the relationship between topology, crystal structure, and various magnetic and electronic characteristics. With the help of external influences in TSM Heusler alloys, it is quite easy to tune the band structure and, consequently, change the density of current carriers and the Hall conductivity. This, in turn, can find its practical application, for example, in the implementation of the quantum anomalous Hall effect in the region of room temperatures.

The study of noncollinear antiferromagnetic Heusler compounds, primarily alloys based on Mn$_3$Z (Z = Sn, Ge, Ga), is of great fundamental and practical interest. In such compounds, the magnetic moments of the Mn atoms are ordered in a kagome-type lattice, which leads to unusual kinetic and magnetic properties. They exhibit a large anomalous Hall effect, a spin Hall effect, and a large spin Nernst effect. Features of the electronic structure and magnetic state lead to the appearance of a large spin-orbital torque, which can be used in antiferromagnetic spintronics for the so-called conversion of the charge current into spin current.

In conclusion, we note that in real Heusler compounds it is not always possible to strictly separate the states considered in this review. Often, HMF, SGS, TSM, and nAFM states can coexist in the same compound. To separate and highlight one or another of them, external influences can be used: temperature, magnetic field, pressure, etc. In addition, when studying the kinetic properties, a large potential is contained in the two-current model of conductivity, briefly considered in this review. Finally, depending on the alloy composition and external parameters, transitions between the considered states, - semiconductor and metallic, nonmagnetic and magnetic phases, are possible. All this opens up further prospects for controlling the electronic and magnetic characteristics of such compounds and their practical application.

**Acknowledgements**

The authors consider it their pleasant duty to thank their colleagues and co-authors M.I. Katsnelson, Yu.N. Skryabin, N.G. Bebenin, E.I. Shreder, A.V. Lukoyanov, Yu.A. Perevozchikova, V.G. Pushin, E.B. Marchenkova for valuable discussions, and A.N. Perevalova, P.S. Korenistov, S.M. Emelyanova, V.V. Chistyakov for help with design of the review.

The research was carried out within the state assignment of Ministry of Science and Higher Education of the Russian Federation (themes "Spin", No. 122021000036-3 and "Quantum", No. 122021000038-7). The research funding from the Ministry of Science and Higher Education of the Russian Federation (Ural Federal University Program of Development within the Priority-2030 Program) is gratefully acknowledged also. The section "Noncollinear antiferromagnets based on Mn$_3$Z Heusler alloys" was prepared with the financial support of the Russian Science Foundation within the framework of research project No. 22-22-00935.